\documentclass[aps,prl,twocolumn,twoside,notitlepage,preprintnumbers,showpacs,
showkeys,byrevtex,floatfix,nofootinbib,amsmath,amssymb]
{revtex4-1}
\usepackage{array,amscd,amsmath,amssymb,graphicx,mathtools, multirow,placeins,slashed,verbatim,subfig,tensor,dsfont}
\usepackage[bookmarksnumbered,bookmarksopen=true,bookmarksopenlevel=1,pdfstartview=FitH]{hyperref}
\usepackage[all]{hypcap}
\usepackage{rotating}
 
\newcolumntype{M}[1]{>{$}{#1}<{$}}

\newcommand{\sst}[1]{{\scriptscriptstyle #1}}

\def\0{{\sst{(0)}}}
\def\1{{\sst{(1)}}}
\def\2{{\sst{(2)}}}
\def\3{{\sst{(3)}}}
\def\4{{\sst{(4)}}}
\def\5{{\sst{(5)}}}
\def\6{{\sst{(6)}}}
\def\7{{\sst{(7)}}}

\newcommand{\be}{\begin{equation}}
\newcommand{\ee}{\end{equation}}
\def\ba{\begin{array}}
\def\ea{\end{array}}

\newcommand{\bea}{\begin{eqnarray}}
\newcommand{\eea}{\end{eqnarray}}

\DeclareMathOperator{\SO}{SO}

\newcommand{\R}{\mathds{R}}
\newcommand{\C}{\mathds{C}}
\newcommand{\Q}{\mathds{H}}
\newcommand{\Oct}{\mathds{O}}
\newcommand{\N}{\mathcal{N}}

\begin{document}

\title{Yang-Mills origin of gravitational symmetries}
\author{A. Anastasiou}
\email[]{alexandros.anastasiou07@imperial.ac.uk}
\affiliation{Theoretical Physics, Blackett Laboratory, Imperial College London, London SW7 2AZ, United Kingdom}
\author{L. Borsten}
\email[]{leron.borsten@imperial.ac.uk}
\affiliation{Theoretical Physics, Blackett Laboratory, Imperial College London, London SW7 2AZ, United Kingdom}
\author{M. J. Duff}
\email[]{m.duff@imperial.ac.uk}
\affiliation{Theoretical Physics, Blackett Laboratory, Imperial College London, London SW7 2AZ, United Kingdom}
\author{L. J. Hughes}
\email[]{leo.hughes07@imperial.ac.uk}
\affiliation{Theoretical Physics, Blackett Laboratory, Imperial College London, London SW7 2AZ, United Kingdom}
\author{S. Nagy}
\email[]{s.nagy11@imperial.ac.uk}
\affiliation{Theoretical Physics, Blackett Laboratory, Imperial College London, London SW7 2AZ, United Kingdom}
\date{\today}

\begin{abstract}

By regarding gravity as the convolution of left and right Yang-Mills theories together with a spectator scalar field in the bi-adjoint representation, we derive in linearised approximation the gravitational symmetries of general covariance, $p$-form gauge invariance, local Lorentz invariance and local supersymmetry from the flat space Yang-Mills symmetries of local gauge invariance and global super-Poincar\'{e}. As a concrete example we 
focus on the new-minimal (12+12) off-shell version of simple four-dimensional supergravity obtained by tensoring the off-shell Yang-Mills multiplets $(4+4,{\cal N}_L=1)$ and $(3+0,{\cal N}_R=0)$. 

\end{abstract}


\preprint{Imperial-TP-2014-MJD-03}

\maketitle



In Einstein's general theory of relativity the requirement that the laws of physics be the same to all observers is embodied in the principle of general covariance: the equations must be invariant under arbitrary coordinate transformations.  Fermions require in addition local Lorentz invariance and many models of interest also incorporate local supersymmetry and local $p$-form gauge invariance.  The purpose of this paper is to derive, in linearised approximation, all these gravitational symmetries starting from those of flat-space Yang-Mills, namely local gauge invariance and global (super)-Poincar\'{e}. 

Early attempts to derive gravity from Yang-Mills were based on gauging spacetime symmetries such as Lorentz, Poincar\'{e} or de Sitter \cite{1956PhRv..101.1597U,1961jMP.....2..212k,1977PhRvL..38..739M,1977NuPhB.129...39C,1979jPhA...12L.205S}. More recently, the AdS/CFT correspondence has provided a different link between gravity and gauge theories \cite{Maldacena:1997re,Witten:1998qj,Gubser:1998bc}.  However, our approach will differ from both in important respects. We appeal to the idea of ``Gravity as the square of Yang-Mills'' by tensoring left and right multiplets with arbitrary non-Abelian gauge groups $G_L$ and $G_R$.  Squaring Yang-Mills is a recurring theme in attempts to understand the quantum theory of gravity  and appears in several different forms: Closed states from products of open states and KLT relations in string theory \cite{kawai:1985xq, Siegel:1988qu, Siegel:1995px},  on-shell $D=10$ Type IIA and IIB  supergravity little-group representations from on-shell $D=10$ super Yang-Mills little-group representations \cite{Green:1987sp}, asymmetric orbifold contructions \cite{Sen:1995ff, Carrasco:2012ca}, gravity anomalies from gauge anomalies \cite{Antoniadis:1992sa}, (super)gravity  scattering amplitudes from those of (super) Yang-Mills  \cite{Bern:2008qj, Bern:2010ue, Bern:2010yg,Bianchi:2008pu,Huang:2012wr,Cachazo:2013iea,Dolan:2013isa,Chiodaroli:2014xia} in various dimensions, (ambi)twistor strings \cite{Hodges:2011wm, Mason:2013sva, Geyer:2014fka} etc.

 In a recent paper \cite{Borsten:2013bp} we addressed the Yang-Mills origin of the {\it global} non-compact supergravity U-dualities $\mathcal{G}$ and their compact subgroups $\mathcal{H}$ by giving a division algebra  $\R,\C,\Q,\Oct$ description of $D=3$ Yang-Mills with $\mathcal{N}=1,2,4,8$. Tensoring left and right on-shell multiplets yields a magic square  $\R\R$, $\C\R$, $\C\C$, $\Q\R$, $\Q\C$, $\Q\Q$, $\Oct\R$, $\Oct\C$, $\Oct\Q$, $\Oct\Oct$ description of $D=3$ supergravity with $\mathcal{N}=2,3,4,5,6,8,9,10,12,16$. This $4 \times4$ square in $D=3$ is the base of a ``magic pyramid'' with a $3 \times 3$ square in $D=4$, a $2 \times 2$ square in $D=6$ and Type II supergravity at the apex in $D=10$ \cite{Anastasiou:2013hba}.  In this paper we focus instead on the Yang-Mills origin of the {\it local} gravitational symmetries of general covariance, local Lorentz invariance, local supersymmetry and $p$-form gauge invariance acting on the classical fields.  
 
 Although much of the squaring literature invokes taking a product of left and right Yang-MiIls fields
 \begin{equation}
 A_\mu(x)(L)\otimes A_\nu(x)(R)
 \label{Siegel}
 \end{equation}
it is hard to find a conventional field theory definition of the product. Where do the gauge indices go? Does it obey the Leibnitz rule
\begin{equation}
\partial_\mu (f \otimes g)=(\partial_\mu f) \otimes g+f \otimes (\partial_\mu  g)
\label{Leibnitz}
\end{equation}
If not, why not?  Since the idea of squaring originates with the $open \otimes open=closed$ property of string theory, one route is to  go back and forth  between asking how string states would behave and then imposing the same rules on the fields. See, for example, Siegel \cite{Siegel:1988qu, Siegel:1995px}. Here we  present a $G_L \times G_R$ product rule within field theory which is valid whether or not there is an underlying string interpretation:
\begin{equation}
 [A_\mu{}^i(L)\star \Phi_{ii'} \star A_\nu{}^{i'}(R)](x)
 \end{equation}
where  $\Phi_{ii'}$ is the ``spectator'' bi-adjoint scalar field  introduced by Hodges \cite{Hodges:2011wm} and Cachazo \emph{et al} \cite{Cachazo:2013iea} and where $\star$  denotes a convolution
\be
[f\star g](x)=\int d^4yf(y)g(x-y).
\ee
 Note $f\star g=g\star f,
 (f\star g)\star h= f\star (g\star h),$ and, importantly obeys
\be
\partial _\mu (f\star g)=(\partial_{\mu} f)\star g=f \star (\partial_{\mu}g)
\ee
and not (\ref{Leibnitz}).

Working with covariant fields rather than physical states favours an off-shell formalism. So although in principle our construction should work in any spacetime dimension $D$ in the purely bosonic case and any $D\leq10$ in the supersymmetric case, we prefer to work with supermultiplets for which the off-shell formalism is known. For concreteness we focus on ${\cal N}=1$ supergravity in $D=4$, obtained by tensoring the $(4+4)$ off-shell ${\cal N}_L=1$ Yang-Mills multiplet $(A_{\mu}{}(L), \chi(L), D(L))$ with the $(3+0)$ off-shell $\N_R=0$ multiplet $A_{\mu}(R)$.  Interestingly enough, this yields the new-minimal formulation of ${\cal N}=1$ supergravity \cite{Sohnius:1981tp} with its 12+12 multiplet  $(e_{\mu}^a,\psi_{\mu}, V_{\mu} , B_{\mu\nu})$, whose transformation properties we now recall, first in superspace and then in components.

The new-minimal formulation of $\N=1$ supergravity is described by a superfield $\varphi_{\mu}$ transforming at linearised level separately under local transformations with chiral parameter $S_\mu$ and real parameter $\phi$ and  under global super-Poincar\'{e} with parameters $a, \lambda, \epsilon$:
\be \label{eq:nmvar}
\delta \varphi_{\mu}=S_\mu+{\bar S}_\mu+{\partial}_{\mu}\phi+\delta_{(a, \lambda, \epsilon)} \varphi_{\mu}
\ee
where $\delta_{(a, \lambda, \epsilon)} F=(aP+\lambda M+\epsilon Q+\bar{\epsilon}\bar{Q})F$ \cite{Cecotti:1987qe, Ferrara:1988qxa}. 

We shall now derive this result by tensoring left and right Yang-Mills multiplets. The left supermultiplet is described by a vector superfield $V^i(L)$ transforming at linearised level separately under local Abelian gauge transformations with parameter $\Lambda^i(L)$, non-Abelian global $G_L$ transformations with parameter $\theta^i(L)$ and global super-Poincar\'{e}:
\be\label{left}
\begin{split}
\delta V^i(L)&=\Lambda^i(L) +\bar \Lambda^i(L)+f^i{}_{jk}V^j(L)\theta^k(L)\\
&\phantom{=}+\delta_{(a, \lambda, \epsilon)} V^i(L).
\end{split}
\ee
Similarly the right Yang-Mills field  $A_{\nu}{}^{i'}{}(R)$ transforms separately under local Abelian gauge transformations with parameter $\sigma^{i'}(R)$, non-Abelian global $G_R$ transformations with parameter $\theta^{i'}(R)$ and global Poincar\'{e}:
\be
\begin{split}
\delta A_\nu{}^{i'}{}(R)&=\partial_\nu\sigma^{i'}(R)+f^{i'}{}_{j'k'}A_\nu{}^{j'}(R)\theta^{k'}(R)\\
&\phantom{=}+\delta_{(a, \lambda)}A_\nu{}^{i'}{}(R).
\end{split}
\ee
The spectator bi-adjoint scalar field transforms under non-Abelian global $G_L \times G_R$  and global Poincar\'{e}:
\be
\delta\Phi_{ii'}=-f^j{}_{ik}\Phi_{ji'}\theta^{k}(L)-f^{j'}{}_{i'k'}\Phi_{ij'}\theta^{k'}(R)+\delta_{a}\Phi_{ii'}.
\ee

The gravitational  symmetries are reproduced here from those of Yang-Mills by invoking the gravity/Yang-Mills dictionary:
\be\label{eq:sfdict}
\begin{split}
 \varphi_{\mu}&= V^i(L)\star \Phi_{ii'}\star A_\mu{}^{i'}{}(R)\\[4pt]
\phi&=V^i(L)\star \Phi_{ii'}\star \sigma^{i'}(R)\\[4pt]
S_\mu&=\Lambda^i(L)\star \Phi_{ii'} \star A_\mu{}^{i'}(R)\\[4pt]
\end{split}
\ee

Noting that the structure constant terms cancel, the variation  
\be
\begin{split}
\delta \varphi_{\mu}&= \delta V^i(L)\star \Phi_{ii'}\star A_\mu{}^{i'}{}(R)\\
&\phantom{=}+V^i(L)\star \delta \Phi_{ii'}\star A_\mu{}^{i'}(R)\\
&\phantom{=}+V^i(L)\star \Phi_{ii'}\star \delta A_\mu{}^{i'}{}(R),\\
\end{split}
\ee
gives
\[
\begin{split}
\delta \varphi_{\mu}&=[\Lambda^i(L) +\bar \Lambda^i(L) +\delta_{(a, \lambda, \epsilon)} V^i(L)]\star \Phi_{ii'}\star A_\mu{}^{i'}{}(R)\\
&\phantom{=}+V^i(L)\star \delta_{a}\Phi_{ii'}\star A_\mu{}^{i'}{}(R)\\
&\phantom{=}+V^i(L)\star \Phi_{ii'}\star[ \partial_\mu\sigma^{i'}(R)+\delta_{(a, \lambda)}A_\nu{}^{i'}{}(R)]\\
&=S_\mu+{\bar S}_\mu+{\partial}_{\mu}\phi+\delta_{(a, \lambda, \epsilon)} \varphi_{\mu}\\
\end{split}
\]
in agreement with \eqref{eq:nmvar}.

To make contact with component formalism we go to Wess-Zumino (WZ) gauge. As usual one applies a supergauge transformation with field dependent chiral parameter $\Lambda^i(L)\vert_{WZ}$ to reduce $V^i(L)$ to the canonical off-shell super-Yang-Mills multiplet,
\be
\begin{split}
V^{i}(L)\vert_{WZ}&= -{ \theta}\sigma^{\mu}{\bar \theta}A_{\mu}{}^{i}(L)\\
&\phantom{=}+i{\theta}^2{\bar \theta}{\bar \chi}^{i}(L)-i{\bar \theta}^2{\theta}\chi^{i}(L)\\&\phantom{=}+\frac{1}{2}{\bar \theta}^2{\theta}^2 D^{i}(L),\\
\end{split}
\ee
 leaving only the standard (Abelian)  gauge transformations $\delta A_{\mu}{}^{i}(L) =\partial_\mu \sigma^{i}(L)$ unfixed.

In this gauge the dictionary \eqref{eq:sfdict} leaves   three  non-zero components,
\be\label{eq:nmmultiplet}
\begin{array}{clllllllll}
Z_{\mu\nu} \equiv h_{\mu\nu}+B_{\mu\nu} &=A_{\mu}{}^{i}(L)&\star& \Phi_{ii'}&\star& A_{\nu}{}^{i'}(R)\\ [4pt]
 \psi_\nu &=\chi^{i}(L)&\star &\Phi_{ii'}&\star& A_{\nu}{}^{i'}(R) \\ [4pt]
V_\nu &=D^{i}(L)&\star &\Phi_{ii'}&\star& A_{\nu}{}^{i'}(R),\\ 
\end{array}
\ee
 which correspond to the conventional new-minimal multiplet. 
 However, we must still check consistency with the parameter dictionary given in \eqref{eq:sfdict}. Applying the chiral supergauge transformation, 
\be
S_\mu\vert_{WZ}  =\Lambda^i(L)\vert_{WZ}\star \Phi_{ii'} \star A_\mu{}^{i'}(R),
\ee
implied by the WZ gauge fixing of $V^i(L)$ consistently  reduces the new-minimal superfield $\varphi_{\mu}$ to  precisely the three components determined by the field dictionary   \eqref{eq:nmmultiplet}. The remaining vector supergauge parameter,
\be
\phi\vert_{WZ}=V^i(L)\vert_{WZ} \star \Phi_{ii'}\star \sigma^{i'}(R)
\ee
 is then comprised of only the required local gauge transformations. Explicitly $\delta \varphi_{\mu}\vert_{WZ} = \partial_{\mu} \phi\vert_{WZ}$ gives
 \be
\begin{array}{clllll}
\delta Z_{\mu\nu}&=\partial_\nu  \alpha_\mu(L)+\partial_\mu  \alpha_\nu(R), \\ 
\delta \psi_\mu &=\partial_\mu\eta,\\ 
 \delta V_\mu &= \partial_\mu \Lambda,
\end{array}
\ee 
where
 \be
\begin{array}{cllllllll}
  \alpha_\mu(L)&=&A_{\mu}{}^{i}(L)&\star &\Phi_{ii'}&\star &\sigma^{i'}(R), \\ [4pt]
    \alpha_\nu(R)&=&\sigma^{i}(L)&\star &\Phi_{ii'}&\star& A_{\nu}{}^{i'}(R), \\ [4pt]
  \eta &=&\chi^{i}(L)&\star& \Phi_{ii'}&\star &\sigma^{i'}(R),\\ [4pt]
  \Lambda&=&D^{i}(L)&\star  &\Phi_{ii'}&\star& \sigma^{i'}(R),
\end{array}
\ee
illustrating how the local gravitational symmetries of general covariance, 2-form gauge invariance, local supersymmetry and local chiral  symmetry follow from those of Yang-Mills.

Similar rules may also be found in \cite{Siegel:1988qu, Siegel:1995px} by applying to fields the product \eqref{Siegel} which is well-defined only when acting on open string states. The issues of gauge indices and the Leibnitz rule are not addressed. By contrast, we insist on a self-contained field-theoretic approach\footnote{Do all theories  obtained by squaring Yang-Mills have a stringy origin? We remain agnostic on this point, noting for example that squaring does not guarantee anomaly freedom.}.

To preserve the WZ gauge the supersymmetry transformation of $V^i(L)$ is accompanied by a field dependent compensating supergauge transformation
\be
 \delta_{\epsilon}^{WZ}V^i(L)=\delta_\epsilon V^i(L)+\Lambda^{i}_{WZ}(L) +\bar \Lambda^{i}_{WZ}(L).
\ee
This leads to a gravitational compensating supergauge transformation  via the dictionary,
\be
\delta_{\epsilon}^{WZ}\varphi_{\mu}=\delta_\epsilon \varphi_{\mu}+S^{WZ}_{\mu} +\bar S^{WZ}_{\mu},
\ee
which consistently preserves the WZ gauge choice imposed on $\varphi_{\mu}$. By making a field redefinition $V_\mu \rightarrow V_\mu+H_\mu$, where $H_\mu = (\star dB)_\mu$,  and shifting the  gauge parameters $\eta \rightarrow \eta+\frac{i}{8}\gamma^{\mu\nu} B_{\mu\nu}\epsilon, \Lambda \rightarrow \Lambda+\bar{\epsilon}\gamma_\lambda\gamma_5\psi^\lambda$ we recover from the dictionary the component supersymmetry variation of \cite{Sohnius:1981tp},
\be
\begin{split}\delta_{\epsilon}^{WZ}Z_{\mu\nu}&=-4i\bar{\epsilon}\gamma_\nu\psi_\mu,\\
\delta_{\epsilon}^{WZ}\psi_\mu&=-\tfrac{i}{4}\sigma^{k\lambda}\epsilon\partial_k g_{\lambda\mu}+\gamma_5\epsilon V_\mu\\&\phantom{=}-\gamma_5\epsilon H_\mu-\tfrac{i}{2}\sigma_{\mu\nu}\gamma_5\epsilon H^\nu,\\
\delta^{WZ}_{\epsilon} V_\mu&=-\bar{\epsilon}\gamma_\mu\sigma^{\kappa\lambda}\gamma_5\partial_\kappa\psi_{\lambda}.
\end{split}
\ee 

New-minimal supergravity also admits an off-shell Lorentz multiplet $(\Omega_{\mu ab}{}^-, \psi_{ab},-2V_{ab}{}^+)$. The definition of this multiplet in terms of new minimal component fields $(e_{\mu}^{a}, \psi_{\mu}, B_{\mu\nu}, V_{\mu})$ can be found in \cite{deRoo:1991at}. It may also be described by a vector superfield $\mathcal{V}^{ab}$ transforming at the linearised level as
\be
\delta \mathcal{V}^{ab}=\Lambda^{ab} +\bar \Lambda^{ab} +\delta_{(a, \lambda, \epsilon)} \mathcal{V}^{ab}.
\ee
This may also be derived by tensoring the left  
Yang-Mills superfield $V^i(L)$ with the right Yang-Mills field strength $F{}^{abi'}(R)$ using the dictionary
\be
\begin{split}
 \mathcal{V}^{ab}&= V^i(L)\star \Phi_{ii'}\star F^{abi'}(R),\\[4pt]
\Lambda^{ab}&=\Lambda^i(L)\star \Phi_{ii'} \star F^{abi'}(R).\\[4pt]
\end{split}
\ee
The corresponding Riemann tensors including torsion terms, as defined in \cite{deRoo:1991at},  are given by
\be
 R_{\mu\nu\rho\sigma}^{+}=-F_{\mu\nu}{}^i(L)\star\Phi_{ii'}\star F_{\rho\sigma}{}^{i'}(R)= R_{\rho\sigma\mu\nu}^{-}. 
\ee
 \vspace{0.01in}
 
\noindent It then follows that  the Yang-Mills equations in Lorenz gauge imply Einstein's equations in De Donder gauge.

Just as convoluting the off-shell Yang-Mills multiplets $(4+4,{\cal N}_L=1)$ and $(3+0,{\cal N}_R=0)$ yields the $12+12$ new-minimal off-shell ${\cal N}=1$ supergravity, so we expect that convoluting the off-shell general multiplet $(8+8,{\cal N}_L=1)$ and $(3+0,{\cal N}_R=0)$ yields the $24+24$ non-minimal off-shell ${\cal N}=1$ supergravity \cite{Breitenlohner:1977jn, Breitenlohner:1976nv} and convoluting $(4+4,{\cal N}_L=1)$ and $(4+4,{\cal N}_R=1)$ yields the $32+32$ minimal off-shell ${\cal N}=2$ supergravity \cite{Fradkin:1979cw, deWit:1979pq, Breitenlohner:1979np, Breitenlohner:1980ej}. The latter would involve Ramond-Ramond bosons from the product of left and right fermions.

Clearly two important improvements would be to generalise our results to the full non-linear transformation rules and to address the issue of dynamics as well as symmetries. 
Dynamics requires gauge-fixing and the inclusion of (anti)ghosts in the dictionary \cite{Siegel:1988qu,Siegel:1995px}. According to \cite{Siegel:1988qu,Siegel:1995px} the  $12+12$ multiplet   splits  with respect to superconformal transformations into an $8+8$ conformal supergravity multiplet plus a $4+4$ conformal tensor multiplet,
\be
\underbrace{\begin{pmatrix}
\mathbf{5+3+1+3}\\
\mathbf{4+2+4+2}\\
\end{pmatrix}}_{\text{new-minimal}}
\rightarrow \underbrace{\begin{pmatrix}
\mathbf{5+3}\\
\mathbf{4+4}\\
\end{pmatrix}}_{\text{conformal}}+\underbrace{\begin{pmatrix}
\mathbf{3+1}\\
\mathbf{2+2}\\
\end{pmatrix}}_{\text{tensor}}
\ee
in terms of  $\SO(3)$ representions. 
Since the left (anti)ghost is a chiral superfield the ghost-antighost sector  gives a compensating $4+4$ chiral (dilaton) multiplet  \cite{Siegel:1988qu,Siegel:1995px}, yielding   old-minimal  $12+12$ supergravity \cite{Stelle:1978ye, Ferrara:1978em}  coupled to a tensor multiplet, which, with the conventional 2-derivative Lagrangian,   correctly corresponds to the on-shell content obtained by tensoring left/right helicity states.

We might speculate that the supergravity $\varphi_{\mu}$, the left Yang-Mills $V^i(L)$, the right Yang-Mills $A_{\mu}{}^{i'}(R)$ and the spectator $\Phi_{ii'}$ live in different worlds with their own Lagrangians. In this case the $n$-point correlation functions would factorize:
\begin{widetext}
\be
\begin{split}
\langle \varphi_{\mu_1}...\varphi_{\mu_n}\rangle
&=\langle V^{i_1}(L)\star \Phi_{{i_1}{i'_1}}\star A_{\mu_1}{}^{i'_1}{}(R)...V^{i_n}(L)\star \Phi_{{i_n}{i'_n}}\star A_{\mu_n}{}^{i'_n}{}(R)\rangle\\
&=\langle V^{i_1}(L)...V^{i_n}(L)\rangle \star \langle \Phi_{{i_1}{i'_1}}...\Phi_{{i_n}{i'_n}} \rangle \star \langle A_{\mu_1}{}^{i'_1}{}(R)...A_{\mu_{n}}{}^{i'_{n}}(R)\rangle\\
\end{split}
\ee
\end{widetext}
(Note that the alternative dictionary $ \Phi^{ii'}\star \varphi_{\mu}= V^i(L)\star  A_\mu{}^{i'}{}(R)$ would impose unacceptably strong constraints on the Yang-Mills fields.) One might then expect to find relations between spin $s$ scattering amplitudes $M(s)$ of the kind discussed in the double-copy literature \cite{Cachazo:2013iea} in the context of BCJ kinematic/color duality \cite{Bern:2008qj, Bern:2010ue, Bern:2010yg,Huang:2012wr}: 
\be
\begin{split}
M(2)&=M(1)M^{-1}(0)M'(1),\\
M(\tfrac{3}{2})&=M(\tfrac{1}{2})M^{-1}(0)M'(1),\\
M(1)&=M(\tfrac{1}{2})M^{-1}(0)M'(\tfrac{1}{2}).\\
\end{split}
\ee

\acknowledgments We are grateful to Biancha Cerchiai, Stanley Deser, Sergio Ferarra, Paul Howe,  Andrew Hodges, Arthur Lipstein, Alessio Marrani, David Skinner, Kelly Stelle and Warren Siegel for useful conversations on squaring Yang-Mills and especially to Zvi Bern for his interest, encouragement and critique of earlier attempts. The work of LB is supported by an Imperial College Junior Research Fellowship. The work of MJD is supported by the STFC under rolling grant ST/G000743/1. AA, LJH and SN are supported by STFC  and EPSRC PhD studentships, respectively.


%

\end{document}